\documentclass[%
 reprint,
 superscriptaddress,
 amsmath,amssymb,
 aps,
]{revtex4-2}
\usepackage{natbib}
\usepackage{braket}
\usepackage{graphicx}
\usepackage{tabularx}
\usepackage{graphicx}
\usepackage{amsmath, amssymb}
\usepackage{bbold}
\usepackage{amsmath}
\usepackage{comment}
\usepackage[normalem]{ulem}
\usepackage[dvipsnames]{xcolor}
\usepackage{qcircuit}

\def\Tr{\mathrm{Tr}}
\def\nq{n_q}

\begin{document}

\preprint{APS/123-QED}

\title{
Quantum algorithms with local particle number conservation: noise effects and error correction
}

\author{Michael Streif}
\affiliation{Quantum Artificial Intelligence Laboratory (QuAIL), NASA Ames Research Center, Moffett Field, CA 94035, USA}
\affiliation{USRA Research Institute for Advanced Computer Science (RIACS), Mountain View, CA 94043, USA}
\affiliation{Data:Lab, Volkswagen Group, Ungererstr. 69, 80805 München, Germany}
\affiliation{University Erlangen-Nürnberg (FAU), Institute of Theoretical Physics, Staudtstr. 7, 91058 Erlangen, Germany}

\author{Martin Leib}
\affiliation{Data:Lab, Volkswagen Group, Ungererstr. 69, 80805 München, Germany}

\author{Filip Wudarski}
\affiliation{Quantum Artificial Intelligence Laboratory (QuAIL), NASA Ames Research Center, Moffett Field, CA 94035, USA}
\affiliation{USRA Research Institute for Advanced Computer Science (RIACS), Mountain View, CA 94043, USA}

\author{Eleanor Rieffel}
\affiliation{Quantum Artificial Intelligence Laboratory (QuAIL), NASA Ames Research Center, Moffett Field, CA 94035, USA}

\author{Zhihui Wang}
\affiliation{Quantum Artificial Intelligence Laboratory (QuAIL), NASA Ames Research Center, Moffett Field, CA 94035, USA}
\affiliation{USRA Research Institute for Advanced Computer Science (RIACS), Mountain View, CA 94043, USA}

\date{\today}

\begin{abstract}
Quantum circuits with local particle number conservation (LPNC) restrict the quantum computation to a subspace of the Hilbert space of the qubit register. In a noiseless or fault-tolerant quantum computation, such quantities are preserved. In the presence of noise, however, the evolution's symmetry could be broken and non-valid states could be sampled at the end of the computation. On the other hand, the restriction to a subspace in the ideal case suggest the possibility of more resource efficient error mitigation techniques for circuits preserving symmetries that are not possible for general circuits. Here, we  analyze the probability of staying in such symmetry-preserved subspaces under noise, providing an exact formula for local depolarizing noise. We apply our findings to benchmark, under depolarizing noise, the symmetry robustness of XY-QAOA, which has local particle number conserving symmetries, and is a special case of the Quantum Alternating Operator Ansatz. 
We also analyze the influence of the choice of encoding the problem on the symmetry robustness of the algorithm and discuss a simple adaption of the bit flip code to correct for symmetry-breaking errors with reduced resources. 
\end{abstract}

\maketitle
\section{Introduction}

Beginning with Shor's algorithm \cite{shor1994algorithms} and Grover's
algorithm \cite{grover1996fast} in the 1990s, dozens of algorithms
\cite{qAlgZoo}
have been found with provable speedups against the best classical algorithms
known, and in some cases against any possible classical algorithm.
For many problems, however, it is as yet unclear whether quantum computing
provides an advantage and, if so, how significant an advantage and by what
means. Even for those for which quantum algorithms are established, the
resources required to realize these gains are far greater than current
or near term quantum hardware support.
For example, for problem sizes of cryptographic interest, Shor's algorithm
requires millions of qubits \cite{gidney2019factor} in stark contrast to
today's state-of-the-art devices with tens of qubits
\cite{barends2016digitized, dicarlo2009demonstration,
debnath2016demonstration}. The recent demonstration \cite{arute2019quantum}
of a noisy intermediate-scale quantum (NISQ) device solving
computational tasks - though artificial ones
- in a matter of seconds which state-of-the-art supercomputers
would require hours, days or even years to solve provides evidence that
even the small noisy devices of today
can be faster than the largest classical computing systems on some problems.
An open question is whether NISQ devices will be able to outperform
classical computations on problems of practical interest, particularly
in areas of broad application such as optimization and quantum
simulation. Whether they can or not, they present
an unprecedented opportunity to explore quantum
algorithms empirically, providing insights into algorithms to be
run on larger scale, fault-tolerant quantum computers of the future.
The practical usefulness of NISQ devices for any of these goals depends
on better understanding the effect of noise on various classes of quantum
algorithms, and of techniques to mitigate errors on NISQ devices.

Algorithms that preserve symmetries have  potential advantages with
respect to performance, robustness, ease of analysis, and use of symmetry-aware
error mitigation techniques. Such algorithms may, however, be
particularly susceptible to noise, which in general will not respect
such symmetries.
Here, we consider one such symmetry, the preservation of particle number (or Hamming weight),
which appears, for example, in certain Quantum Alternating Operator Ansatz (QAOA) algorithms,  and in Quantum Variational eigensolver algorithms \cite{McClean16, Ryabinkin19,Seki20,Gard20}.

We examine the effect of noise, and potential error mitigation approaches,
on general quantum circuits that preserve particle number, and look
specifically at the effects of noise and error mitigation approaches
for XY-QAOA, a special case of
the Quantum Alternating Operator Ansatz (QAOA) \cite{hadfield2019quantum}
itself a generalization of the framework used in
the Quantum Approximate Optimization Algorithm (QAOA) \cite{farhi2014quantum}.
The original X-QAOA was designed to find approximate solutions
to unconstrained optimization problems. Most industry-relevant optimization
problems, however, include constraints.
XY-QAOA uses an XY-mixer to preserve particle number. Recent numerical
simulations \cite{wang2020x}
confirmed intuitions that, at least in the noiseless case,
XY-QAOA substantially outperforms X-QAOA for problems
with the appropriate symmetries.

The main contributions of this work are:
\begin{itemize}
\item An analytically-derived exact formula for the probability of
preserving particle number under locally homogeneous noise such as that
of a depolarizing channel for any circuit that preserves particle number
in the noiseless case.
\item Calculation of the probability of preserving particle number
as a function of depth and error rate for XY-QAOA, with specific
applications to QAOA applied to the Max-$\kappa$-Colorable-Subgraph problem,
with the coloring of each vertex mapped to a particle number $1$ subspace.
\item Comparison of alternative mappings of the Max-$\kappa$-Colorable-Subgraph
problem to higher particle number subspaces and their relative robustness.
\item Demonstration of a symmetry-aware error mitigation scheme for particle number preserving algorithms that
uses fewer resources than generic quantum error correction.
\item Open questions and research directions for NISQ algorithms
that preserve symmetries, both particle number symmetries and more general
symmetries.
\end{itemize}

This paper is structured as follows. In Sec.~\ref{sec:probability}, we define a class of quantum circuits with LPNC and calculate the probability of upholding the symmetry during a computation in the presence of local depolarizing noise. In Sec.~\ref{sec:XYPerformance}, we benchmark the robustness of XY-QAOA applied to the Max-$\kappa$-Colorable-Subgraph problem under the influence of noise. In Sec.~\ref{sec:embeddingsanderrormitigation}, we analyze how the choice of encoding the problem affects the robustness to noise and benchmark an adaption of the bit flip code to correct for symmetry-breaking errors. In Sec.~\ref{sec:conclusion}, we conclude and give an outlook. 

\section{Quantum algorithms with particle number symmetries under noise}
\label{sec:probability}
In this first section, we analyze quantum algorithms with conserved particle numbers under noise and give an analytical expression 
for the probably of leaving a fixed particle number subspace. 

We assume a system composed of $n$ subsystems with $\kappa$ qubits each. We initialize each system with a fixed particle number $N$, that is each system has $N$ qubits in the $\ket{1}$ state and $\kappa-N$ qubits in the $\ket{0}$ state. For example, for the $N=1$ particle number subspace, each system has only a single particle.

The subsystem states with particle number $N$ span a subspace, which we refer to as \textit{particle number subspace} $\mathcal{H}_\mathrm{N}$. The tensor product of all $n$ subspaces defines the \textit{feasible subspace} on all $n$ subsystems, $\mathcal{H}_\mathrm{feas}=\mathcal{H}_\mathrm{N}^{(1)}\otimes\mathcal{H}_\mathrm{N}^{(2)}\otimes\dots\otimes\mathcal{H}_\mathrm{N}^{(n)}$. We moreover define a set of one- and two-qubit unitaries $G$ 
such that each element $g\in G$ upholds the local particle number conservation (LPNC) in each subsystem, that is 
\begin{align}
    g N_i = N_i g\hspace{3mm}\forall\phantom{.}g\phantom{.}\mathrm{in}\phantom{.}G\phantom{.}\forall\phantom{.}i\,,
\end{align}
with the particle number operator
\begin{align}
    N_i=\sum_k^{\dim \mathcal{H}_N} \frac{1}{2}(\mathbb{1}^k-\sigma_z^{(i_k)})
\end{align} 
with $i_k$ the $k$-th qubit of subsystem $i$
counting how man qubits are in the $\ket{1}$ state in subsystem $i$, i.e. $g|_{\mathcal{H}_\mathrm{feas} }:\mathcal{H}_\mathrm{feas}\rightarrow\mathcal{H}_\mathrm{feas}\phantom{.}$. We also introduce a local depolarizing noise channel $\mathcal{E}(\rho):\rho \rightarrow \sum_i K_i\rho K_i^\dagger$ on each qubit by defining the Kraus operators 

\begin{align}
   K_1&= \sqrt{1-\eta}\mathbb{1},\hspace{0.3cm} &K_2&= \sqrt{\frac{\eta}{3}}\sigma_x,\hspace{0.3cm}   \nonumber
\\ K_3&=\sqrt{\frac{\eta}{3}}\sigma_y,\hspace{0.3cm} &K_4&=\sqrt{\frac{\eta}{3}}\sigma_z,
\label{eq:depol}
\end{align}
where $\mathbb{1}$ and $\sigma_{x,y,z}$ are the identity and the three Pauli matrices respectively and $\eta$ is the probability that one of the errors occurs. This channel describes the symmetric decay of a single qubit into the center of the Bloch sphere. Note that, when $\eta=0.75$, the output of this channel is a fully mixed state. 
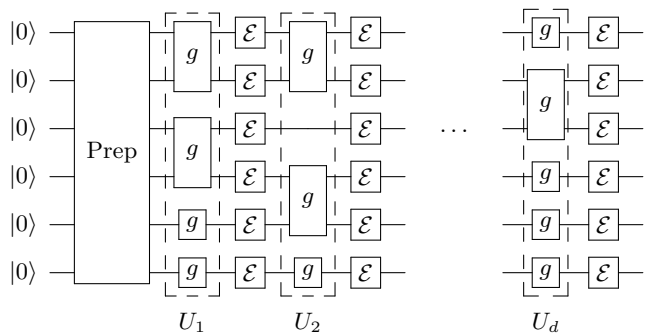
\begin{figure}[t!]
\centering
\leavevmode 
\Qcircuit @C=1.0em @R=0.7em {& & & & & & & & & & &\\
\lstick{\ket{0}} & \multigate{5}{\mathrm{Prep}}  & \multigate{1}{g} & \gate{\mathcal{E}} & \multigate{1}{g} & \gate{\mathcal{E}} &  \qw && &  & & \gate{g} &  \gate{\mathcal{E}}& \qw \\
\lstick{\ket{0}} & \ghost{\mathrm{Prep}} & \ghost{g} & \gate{\mathcal{E}} & \ghost{g} & \gate{\mathcal{E}} &\qw &  && & &   \multigate{1}{g}&  \gate{\mathcal{E}}& \qw\\
\lstick{\ket{0}} & \ghost{\mathrm{Prep}} & \multigate{1}{g} & \gate{\mathcal{E}}& \qw & \gate{\mathcal{E}} &\qw  &&{\dots} & & &   \ghost{g} &  \gate{\mathcal{E}}& \qw \\
\lstick{\ket{0}} & \ghost{\mathrm{Prep}} & \ghost{g} & \gate{\mathcal{E}} & \multigate{1}{g} & \gate{\mathcal{E}} &\qw & & & & &   \gate{g} &  \gate{\mathcal{E}}& \qw \\
\lstick{\ket{0}} & \ghost{\mathrm{Prep}} & \gate{g} & \gate{\mathcal{E}} & \ghost{g} & \gate{\mathcal{E}} & \qw& & & & &    \gate{g} &  \gate{\mathcal{E}}& \qw  \\
\lstick{\ket{0}} & \ghost{\mathrm{Prep}} & \gate{g} & \gate{\mathcal{E}} & \gate{g}& \gate{\mathcal{E}} &\qw & & &  & &  \gate{g}&  \gate{\mathcal{E}} & \qw \\
& & & & & & & & & & &\\
& & \mbox{\hspace{0mm}$U_1$} 
\gategroup{2}{3}{7}{3}{.7em}{--}&&
\mbox{\hspace{0mm}$U_2$} 
\gategroup{2}{5}{7}{5}{.7em}{--}&&&&&&&\mbox{\hspace{0mm}$U_d$} 
\gategroup{2}{12}{7}{12}{.7em}{--}} 
\caption{A circuit illustrating our setup. After initializing the system in a state of $\mathcal{H}_\mathrm{feas}$, randomly drawn one- and two-qubit unitaries from $G$ are applied (note that each g can be a different operation, and we omitted distinct labels for brevity of the notation), followed by local depolarizing channels $\mathcal{E}$.}
\label{fig:circuit}
\end{figure}
In the following, without causing confusion, the expression $\mathcal{E}(\rho)$ on a multi-qubit state $\rho$ is also used to describe the action of the local noise channel on all qubits in $\rho$.

To mimic a quantum circuit with LPNC under noise, we first initialize the system in a state $\rho_0\in\mathcal{B}(\mathcal{H}_\mathrm{feas})$ and subsequently apply a set of one- and two-qubit unitaries drawn from $G$. After each layer of gates, or maximal set of consecutive gates which act on distinct sets of qubits, denoted by $U_l$, we apply the depolarizing channel $\mathcal{E}$ to all qubits, as pictured in Fig.~\ref{fig:circuit}. We thus assume that all operations, including idle times of qubits, are subject to the same amount of noise.

To calculate the probability to escape the feasible subspace $\mathcal{H}_\mathrm{feas}$, we define the projection operator onto the feasible subspace $\mathcal{P}_\mathrm{feas}=\mathcal{P}_\mathrm{N}^{(1)}\otimes\mathcal{P}_\mathrm{N}^{(2)}\otimes\dots\otimes\mathcal{P}_\mathrm{N}^{(n)}$ with $\mathcal{P}_{\mathrm{N}}^{(i)}$ the projector on the particle number subspace with $N$ particles on subsystem $i$. The expectation value $p_\mathrm{feas}=\braket{\mathcal{P}_{\mathrm{feas}}}$ w.r.t. the output state of the quantum computation then is the probability of sampling a valid bitstring. 

By construction, gates from $G$ map states with particle number $N$ to states with the same particle number. Thus, in the absence of noise, the whole circuit preserves the particle number at any point of the computation, and there is no population transfer between different  subspaces, i.e. $p_\mathrm{feas}=1$. 
This changes in the presence of noise, where the expectation value includes noise channels after each layer of the circuit,
\begin{align}
    p_\mathrm{feas}&=\mathrm{Tr}\left[\mathcal{P}_{\mathrm{feas}}  {\cal E}(U_{d}{\cal E}(\cdots{\cal E}(U_1{\cal E}(\rho_0)U^\dag_1) \cdots) U^\dag_{d}) \right]\nonumber\\
    &=\mathrm{Tr}\left[\rho_0 U^\dag_1 {\cal E}(\cdots{\cal E}(U^\dag_d{\cal E}(\mathcal{P}_{\mathrm{feas}})U_d) \cdots)U_1 \right]\,,
    \label{eq:expectationvalue}
\end{align}
where we used that all Kraus operators of the depolarizing channel are Hermitian.
Without loss of generality, we assume in the following that all subsystems have the same size and are affected by the same amount of noise, i.e. have equal value of $\eta$ for all qubits. As a starting point, we exploit the locality of the defined noise channel to write
\begin{align}
{\cal E}(\mathcal{P}_{\mathrm{feas}})&={\cal E}(\mathcal{P}_\mathrm{N}^{(1)}\otimes\mathcal{P}_\mathrm{N}^{(2)}\otimes\dots\otimes\mathcal{P}_\mathrm{N}^{(n)})\\
&={\cal E}(\mathcal{P}_\mathrm{N}^{(1)})\otimes {\cal E}(\mathcal{P}_\mathrm{N}^{(2)})\otimes\dots\otimes{\cal E}(\mathcal{P}_\mathrm{N}^{(n)}).
\end{align}
We note that the Kraus operator $K_4$, cf. Eq.~(\ref{eq:depol}) does not change the population distribution of a qubit and thus commutes with all particle number projectors and thus can be neglected. The Kraus operators $K_2$ and $K_3$ cause local bit flips and thus generate projection operators onto different particle numbers. As all states with a particle number $j$ require the same number of spin flips to reach a state with particle number $j'$, we can write the equation above as a weighted sum of projection operators,
\begin{align}
{\cal E}(\mathcal{P}_{\mathrm{feas}})=\sum_{j=0}^\kappa c_j {\cal P}_{j}^{(1)}\otimes\sum_{j=0}^\kappa c_j {\cal P}_{j}^{(2)}\otimes\dots\otimes\sum_{j=0}^\kappa c_j {\cal P}_{j}^{(n)}\,,
\label{eq:projectionsuperposition_a}
\end{align}
with $c_j$ the weight of the projectors $\mathcal{P}^{(i)}_j\forall i$. The next step in our calculation is the application of the first layer of the circuit onto this sum of projection operators, i.e. $U_1^\dagger{\cal E}(\mathcal{P}_{\mathrm{feas}})U_1$. As we designed the unitaries to respect the LPNC, we have $U^\dag_l  {\cal E}(\mathcal{P}_{\mathrm{feas}})U_l={\cal E}(\mathcal{P}_{\mathrm{feas}}$), that is, ${\cal E}(\mathcal{P}_{\mathrm{feas}})$ stays invariant under the unitaries in $G$. This behaviour remains the same for a cascade of $k$ applications of noise channels and unitaries, 
\begin{align}
&U^\dag_1  {\cal E}(U^\dag_2{\cal E}(\cdots{\cal E}(U^\dag_d{\cal E}(\mathcal{P}_{\mathrm{feas}}^i)U_d) \cdots) U_2)U_1\nonumber\\
=&\sum_{j=0}^\kappa c_j^{(k)} {\cal P}_{j}^{(1)}\otimes\sum_{j=0}^\kappa c_j^{(k)} {\cal P}_{j}^{(2)}\otimes\dots\otimes\sum_{j=0}^\kappa c_j^{(k)} {\cal P}_{j}^{(N)}\,.
\label{eq:projectionsuperposition_b}
\end{align}
Thus, for $k=d$ layers, the expectation value, Eq.~(\ref{eq:expectationvalue}), reads
\begin{align}
\braket{\mathcal{P}_{\mathrm{feas}}}
&=\Tr\left[\left(c_N^{(d)}\mathcal{P}_{\mathrm{N}}^{(1)}\otimes c_N^{(d)}\mathcal{P}_{\mathrm{N}}^{(2)}\otimes \dots c_N^{(d)}\mathcal{P}_{\mathrm{N}}^{(N)}\right)\rho_0\right]\nonumber\\
&=\prod^n c_\mathrm{N}^{(d)}=\left(c_\mathrm{N}^{(d)}\right)^n\,,
\label{eq:probability_feas}
\end{align}
where we used that $\rho_0\in\mathcal{B}(\mathcal{H}_\mathrm{feas})$ with particle number $N$ and that states with other particle number are mutually orthogonal. We observe that this expression is independent of the unitaries $U_l$ and that the exact coefficient $c_\mathrm{N}^{(d)}$ is determined solely by the noise level $\eta$, the circuit depth $d$ and the particle number $N$. The calculation of the coefficient $c_\mathrm{N}^{(d)}$ is a combinatorial task and can be reformulated as the simple question: starting with a classical bitstring of length $\kappa$ with $N$ bits showing up and $\kappa-N$ bits showing down, how likely is it to find a bitstring with $N$ bits showing up after $d$ independent symmetric random processes on each bit? In this picture, each random process is the classical representation of the local quantum error channel and has certain probability to flip the bit. To answer this question, we first give the probability that a single bit is found in its initial state after $d$ random processes,
\begin{align}
    p_{0/1\rightarrow 0/1}(d)=\sum_{k\in \mathbb{N}}^{\left \lfloor{ d/2}\right \rfloor} \binom{d}{d-2k}\tilde{\eta}^{d-2k}(1-\tilde{\eta})^{2k}\,.
\end{align}
In this expression, $\tilde{\eta}=(1-2/3\eta)$ 
only depends on the error rates from Kraus operators $K_2$ and $K_3$. With this probability we count all possibilities to find a bitstring with $N$ bits showing up,
\begin{align}
    c_\mathrm{N}^{(d)}=\sum_{j\in\mathbb{N}}^{M}    \binom{N}{j}\binom{\kappa-N}{j}p_{0/1\rightarrow 0/1}^{\kappa-2j}(1-p_{0/1\rightarrow 0/1})^{2j}.
    \label{eq:probability_N}
\end{align}
where $M =\min\{N, \kappa - N\}$.
For a system composed of $n$ subsystems, the probability to stay in the feasible subspace, $p_\mathrm{feas}=\left(c_\mathrm{N}^{(d)}\right)^n$, only depends on the number of subsystems $n$, the particle number $N$, the noise level $\eta$ and the circuit depth $d$. Evidently, if $d>1$ and $\eta\neq 0$, $c_\mathrm{N}^{(d)}<1$ and $p_\mathrm{feas}$ decays exponentially in the number of subsystems. 

\begin{figure*}[t!]
    \centering
    \begin{minipage}{0.5\textwidth}
        \centering
        \includegraphics[width=1\textwidth]{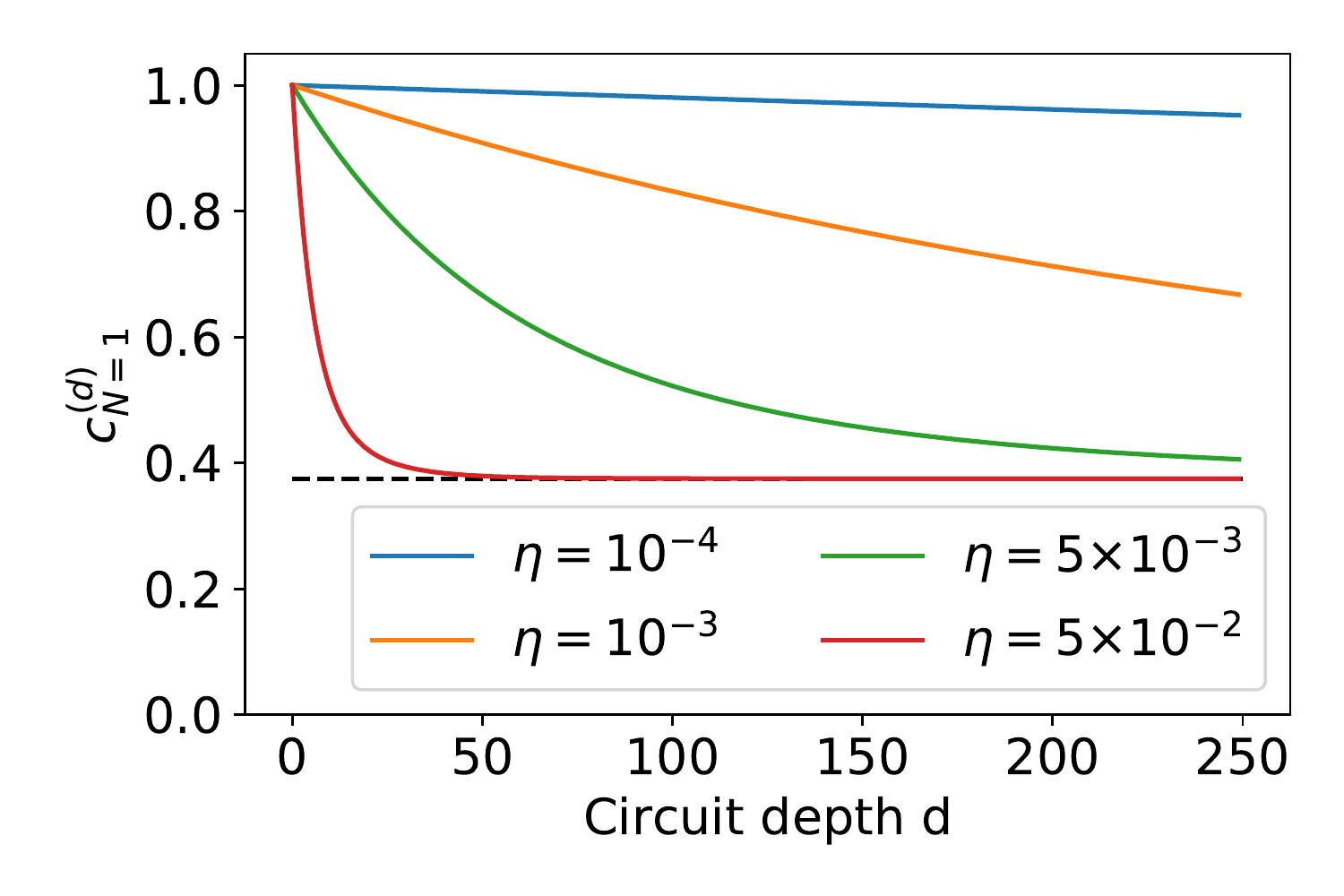}\llap{
  \parbox[b]{6.0in}{(a)\\\rule{0ex}{2.2in}
  }}
    \end{minipage}\hfill
    \begin{minipage}{0.5\textwidth}
        \centering
        \includegraphics[width=1\textwidth]{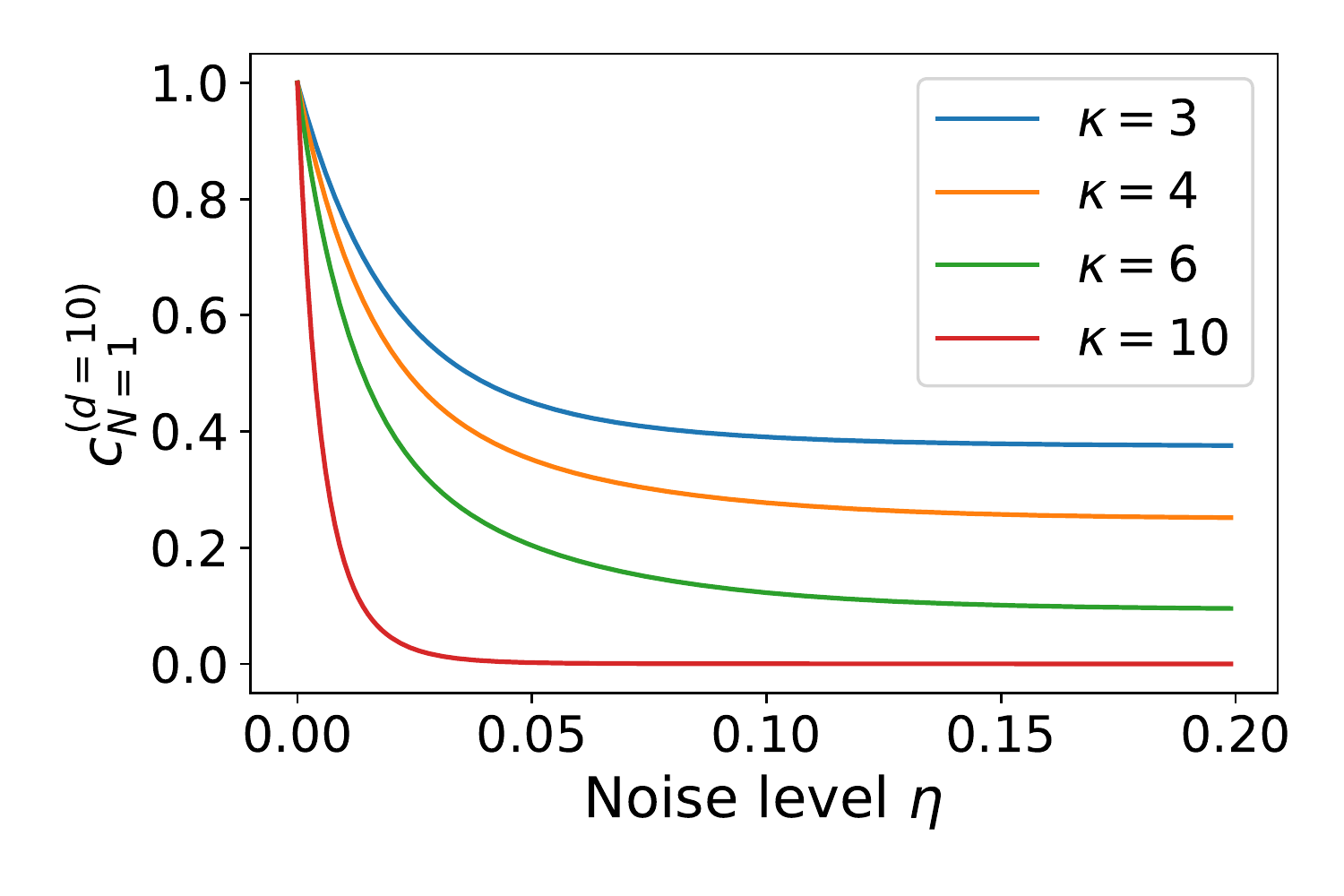}\llap{ \parbox[b]{6.0in}{(b)\\\rule{0ex}{2.2in}
  }}
    \end{minipage}
\caption{(a) The probability of staying in the $N=1$-subspace of $\kappa=3$ qubits for various noise levels as function of the circuit depth. The dashed line shows the probability w.r.t. to a completely mixed state, $\rho_\mathrm{mixed}=\mathbb{1}/2^3$. (b) The monotonically decreasing probability of staying in the $N=1$-subspace as a function of the level of noise $\eta$, shown for four different numbers of colors  $\kappa$ for a circuit of depth $d=10$.}
\label{fig:decay}
\end{figure*}

Fig.~\ref{fig:decay} shows $c_\mathrm{N}^{(d)}$ as function of the circuit depth $d$ for $N=1$ and $\kappa=3$ for various noise levels. The horizontal line shows the expectation value w.r.t. to a completely mixed state, $\mathrm{Tr}\left(\mathbb{1}/2^\kappa \mathcal{P}_{\mathrm{N}}\right)$.
The probability of staying in the feasible subspace monotonically decreases with noise, as shown in Fig.~\ref{fig:decay}(b).

In the following section, we apply the above theory and study the behavior of $p_\mathrm{feas}$ for an example applications in the NISQ era.

\section{Quantifying the robustness of XY-QAOA under depolarizing noise}

\label{sec:XYPerformance}
In this section, we use our findings to analyze the robustness of 
the Quantum Alternating Operator Ansatz,~\cite{hadfield2019quantum}, in particular XY-QAOA~\cite{wang2020x},
an adaption of the  Quantum Approximate Optimization Algorithm (QAOA)~\cite{farhi2014quantum} to solve classical optimization problems subject to hard constraints, under the influence of noise. 

\subsection{Review of the Quantum Alternating Operator Ansatz (QAOA)}

The Quantum Approximate Optimization Algorithm (QAOA) \cite{farhi2014quantum}, and the extension of its framework to the general class of Quantum Alternating Operator Ans\"atze, are leading quantum metaheuristics for exact or approximate solutions of classical optimization problems. Low depth versions of these algorithms are suitable for NISQ devices. While in some  cases good parameters can be found analytically \cite{jiang2017near}, or through decomposition techniques that make use of light or causal cones to obtain analytical formulas for the parameters that can be optimized numerically \cite{farhi2014quantum, wang2018quantum,Streif_2020,zhou2020quantum,bravyi2019classical}, variational approaches remain popular.

Early experiments on real quantum devices have been realized \cite{otterbach2017unsupervised, arute2020quantum}. Despite all this progress, only little is known about its performance in comparison to classical algorithms or its robustness against noise \cite{zhou2020quantum, streif2019comparison, hastings2019classical, Marshall_2020, xue2019effects, streif2020forbidden,2020arXiv201103403S}.

In QAOA, the first step is to map the cost function to a classical spin Hamiltonian commonly referred to as problem Hamiltonian $H_P$. This mapping has to be chosen such that the eigenstates of the Hamiltonian and their corresponding energies are in direct relationship to the solutions and their classical costs of the classical problem respectively.

The problem Hamiltonian generates, together with a second Hamiltonian, often $H_\mathrm{X}=\sum_i\sigma_x^{(i)}$, the mixing and phase separation unitaries, $U_{\mathrm{M}}(\beta_i) = \exp[{-i\beta_i H_\mathrm{M}}]$ and $U_{\mathrm{PS}}(\gamma_i)=\exp[{-i\gamma_i H_\mathrm{P}]}$ respectively, which define the QAOA ansatz state 
\begin{align}
    \left|\Psi(\{\beta_i, \gamma_i\})\right\rangle = \prod_i^{p} U_\mathrm{M}(\beta_i)U_\mathrm{P}(\gamma_i) \left|+\right\rangle^{\otimes n}\,,
\end{align}
with $p$ the number of iterations of both unitaries, or QAOA blocks, and the initial state, $\left|+\right\rangle^{\otimes n}=\bigotimes_i^n 1/\sqrt{2}\left(\ket{0}_i+\ket{1}_i\right)$, the superposition of all computational states. 
The variational parameters $\{\beta_i,\gamma_i\}$ are then optimized aiming to minimize the expected energy of the problem Hamiltonian,
\begin{align}
    E_g = \min_{\left\{\beta_i, \gamma_i\right\}} \left\langle \Psi(\{\beta_i, \gamma_i\})| H_{\mathrm{P}}| \Psi(\{\beta_i, \gamma_i\})\right\rangle \,.
\end{align}
Most optimization problems are however subject to constraints, which shrink the state space of valid solutions. A common approach to include constraints into quantum algorithms, e.g. quantum annealing  or QAOA \cite{perdomo2012finding}, is to penalize all states not fulfilling the constraints by adding an energy term
to the problem Hamiltonian.

In this setup however, the algorithm still accesses the whole state space, including non-valid solutions.

In \cite{hadfield2019quantum, wang2020x}, an alternative approach was introduced. Instead of penalizing the constraint-violating states in the cost Hamiltonian, the mixing Hamiltonian $H_M$ in the QAOA ansatz state was adapted to respect the underlying symmetry of the constraints and thereby restricts the algorithm's evolution to valid states only. 
Generally, the feasible subspace, the subspace of valid states, while still exponentially large, is exponentially smaller than the full search space.
For example, to preserve the particle number on qubits in a subset $S$, a possible choice of the mixing Hamiltonian is 
\begin{align}
    H_\mathrm{XY}&=\sum_{i,j\in S} H_\mathrm{XY}^{i,j} \\
    H_\mathrm{XY}^{i,j}&=
    \frac{1}{2}\left(\sigma_x^{(i)}\sigma_x^{(j)}+\sigma_y^{(i)}\sigma_y^{(j)}\right)
    \,.
    \label{eq:mixing}
\end{align}
The unitary generated by this mixing Hamiltonian, $U_{\mathrm{M}}(\beta)
\rightarrow U_{\mathrm{XY}}(\beta) = \exp[{-i\beta H_\mathrm{XY}}]$, then drives only transitions between states with same particle number. A direct advantage is that the computation is restricted to valid states only, which significantly shrinks the state space the algorithm has access to. In \cite{wang2020x}, it was numerically shown that this approach outperforms the approach with energy penalties. 

However, in the presence of noise, leaving the feasible subspace could have drastic consequences for the performance of the algorithm. 

\subsection{Quantifying robustness under depolarizing noise}
\label{sec:performance_under_noise}
\begin{figure}[t!]
  \includegraphics[width=\linewidth]{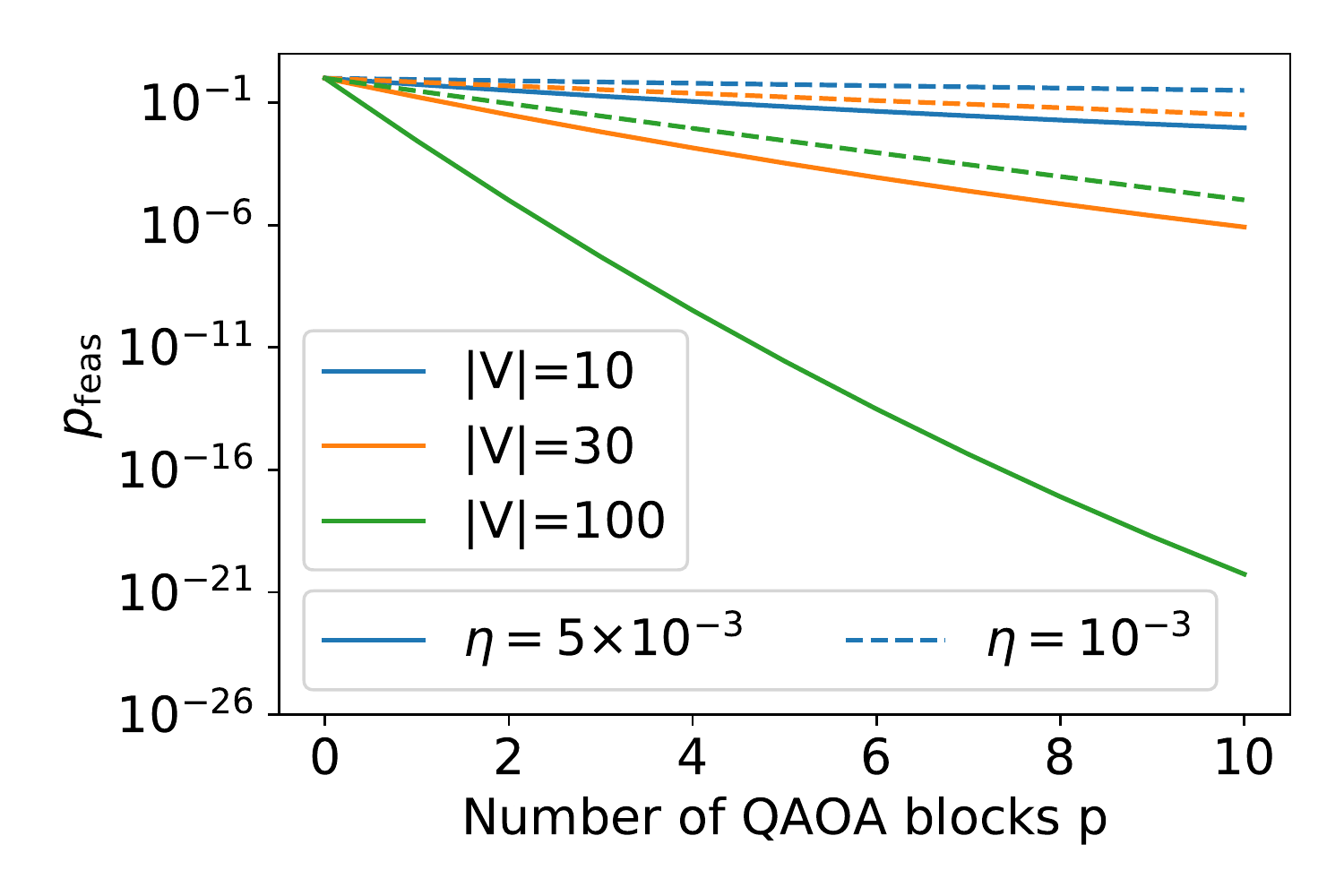}
  \caption{The probability to stay in the feasible subspace for XY-QAOA applied to the Max-$\kappa$-Colorable-Subgraph problem with one-hot encoding and $\kappa=3$ colors for a 3-regular graph with $|V|=\{10,30,100\}$ logical vertices as function of the number of QAOA blocks $p$,
  assuming qubits are fully-connected.
  }
  \label{fig:XY}
\end{figure}

We here use the example from \cite{wang2020x} and analyze the robustness of XY-QAOA applied to the Max-$\kappa$-Colorable-Subgraph problem of a graph $G=(V,E)$ with vertices $V$ and edges $E$ in the presence of noise.  
For simplicity in analysis, we consider 3-regular graphs $G$. 
This problem asks to maximize the number of edges in a correctly vertex-colored subgraph. For a precise definition of the problem, we refer to \cite{wang2020x}. For $\kappa=2$ colors, the internal qubit states $\ket{0}$ and $\ket{1}$ can be used to represent the different colors. 
For more colors however, more than a single qubit is required. Thus, a logical vertex $v\in V$ of the graph has to be encoded non-locally in several qubits of the qubit register. 
A common encoding strategy is the one-hot encoding \cite{okada2019efficient}, where each color is represented by one of the particle number $N=1$ states of $\kappa$ qubits. For the example of three colors, the three states, $\ket{100}$, $\ket{010}$ and $\ket{001}$ could encode the three different colors. The classical cost function encoded in the space spanned by the above states, counts the number of edges that connects two vertices of the same color,
\begin{align}
    f_\mathrm{C} = \sum_j^\kappa \sum_{(v,v')\in E}x_{v,j}x_{v',j}\,,
    \label{eq:cost_one_hot}
\end{align}
with $m=|E|$ the number of edges in the graph $G$.
The corresponding problem Hamiltonian can be formulated by promoting the binary variables to spin operators, $x_{a,b}\rightarrow(\mathbb{1}-\sigma_z^{(a,b)})/2$,
\begin{align}
    H_\mathrm{P}=
    \frac{1}{4}\sum_{j=1}^\kappa \sum_{(v,v')\in E}\sigma_z^{(v,j)}\sigma_z^{(v',j)}\,.
    \label{eq:one_hot_encoding}
\end{align}
Here, we used that the sum over all 1-local terms, $\sum_i\sigma_z^{(v,j)}$ is proportional to the particle number operator $N_v$ and hence constant in the feasible subspace. 

The mixing Hamiltonian Eq.~(\ref{eq:mixing}) is used to drive transitions between all states with single particle of each logical vertex. By construction, both the mixing unitaries, $U_\mathrm{XY}(\beta_i)$, and the phase separation unitaries, $U_\mathrm{P}(\gamma_i)$, preserve the particle number of each vertex-subsystem and satisfy the requirements we made for the unitaries in $G$, cf.  Sec.~\ref{sec:probability}. As this is true for any choice of variational parameters $\{\gamma_i,\beta_i\}$, the probability of staying in the feasible subspace does not depend on the value of the parameters. Thus, we ignore the variational character of QAOA for the following analyses.

We moreover note that the approximation ratio and the ground state population are trivially upper bounded by the probability of staying in the feasible subspace, as all non-feasible states have ground state probability and approximation ratio equal to zero.  
While the actual performance measured by the approximation ratio or ground state probability relies heavily on the parameter updating techniques, we focus on the symmetry aspect in the operators which is parameter-agnostic.
In the rest of the paper we evaluate only the probability of an output state to be in the feasible subspace and will refer this quantity as the robustness unless otherwise stated.

To apply theories developed in Sec.\ref{sec:probability} and use Eq.~(\ref{eq:probability_feas}), we have to quantify the circuit depth required to implement the mixing and problem unitaries, $U_\mathrm{M}$ and $U_\mathrm{P}$. The circuit depth however heavily depends on the topology of the hardware graph. As different technologies, such as ion traps or superconducting qubits, differ in their topology design in NISQ era, we study two different qubit connectivity scenario in the following subsections: (1) qubits are fully-connected and (2) qubits form a 2D grid.  
Moreover, we assume that all gates we define in our further analysis are available as native gates
and do not require further compilation into different gate sets, which would further increase the circuit depth.

\subsubsection{Quantifying robustness on fully-connected hardware}
\label{sec:xyqaoa-fully-connected}
On fully-connected quantum computers, we can carry out all interactions without any overhead of routing the qubits on the hardware.
The mixing Hamiltonian constitutes of interactions between all pairs of qubits representing a logical vertex. 
The corresponding unitary can be approximated by any ordered product of two-qubit unitaries on these pairs, $\exp[-i\beta H_\mathrm{XY}^{i,j}]$ \cite{wang2020x}.

For $\kappa$ colors, the unitary can be carried out with circuit depth $d_\mathrm{XY}=\kappa-1$ if $\kappa$ is even and $d_\mathrm{XY}=\kappa$ if $\kappa$ is odd \cite{ehrenfeucht1984new}. The problem unitaries, generated by the Hamiltonian in Eq.~(\ref{eq:one_hot_encoding}), include two-qubit unitaries between all $\kappa$ qubits representing the same color for each connection $e\in E$ in the logical graph. Thus, we have to carry out a single two-qubit unitary between all color pairs for each connection. On a fully-connected quantum computer, this can be parallelized, resulting in circuit depth $d_\mathrm{P}=(k+1)$ for a $k$-regular graph if $k$ is odd and $d_\mathrm{P}=k$ if  $k$  is even \cite{ehrenfeucht1984new}. Thus, in total, we can realize the circuit in depth $d_\mathrm{QAOA}=p(d_\mathrm{P}+d_\mathrm{XY})$ for $p$ QAOA blocks on a fully-connected quantum computer. As in the previous section, we apply noise channels after each layer in the circuit and calculate the probability of staying in the feasible subspace with Eqs.~(\ref{eq:probability_N}) and (\ref{eq:probability_feas}). In Fig.~\ref{fig:XY}, we plot the probability of staying in the feasible subspace for $\kappa=3$ colors and $k=3$-regular graphs with $|V|=\{10,30,100\}$ vertices for various number of QAOA blocks and noise levels. 

\subsubsection{Quantifying robustness on 2D grids}
\label{sec:routing}
\begin{figure*}[t!]
    \centering
    \begin{minipage}{0.5\textwidth}
        \centering
        \includegraphics[width=1\textwidth]{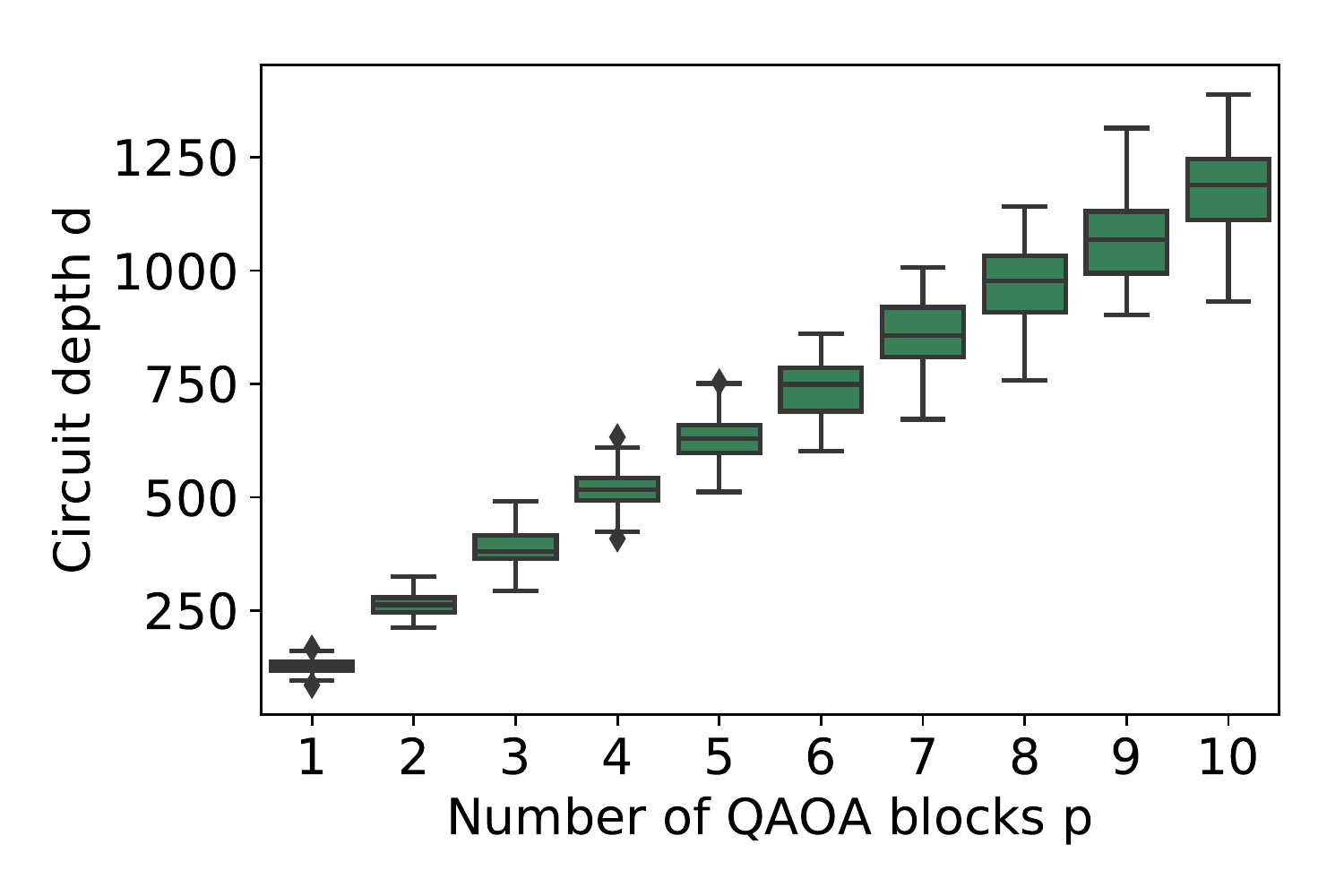}\llap{
  \parbox[b]{6.0in}{(a)\\\rule{0ex}{2.2in}
  }}
    \end{minipage}\hfill
    \begin{minipage}{0.5\textwidth}
        \centering
        \includegraphics[width=1\textwidth]{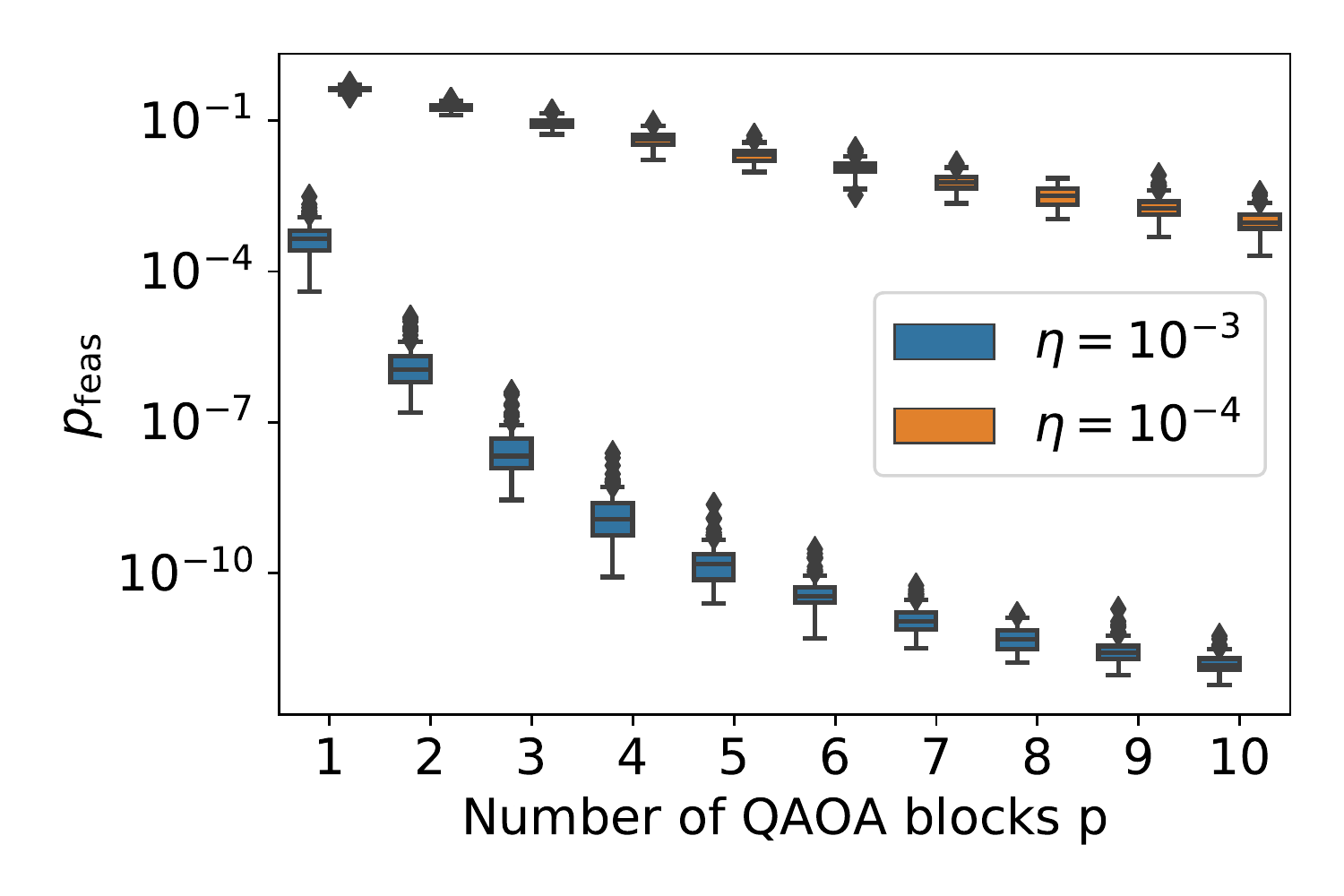}\llap{ \parbox[b]{6.0in}{(b)\\\rule{0ex}{2.2in}
  }}
    \end{minipage}
\caption{Results for the Max-3-Colorable-Subgraph problem with one-hot encoding for 3-regular graphs with $|V|=30$ logical vertices. (a) The circuit depth of the XY-QAOA circuit for various number of blocks $p$ when routing the circuit onto a 2D grid of size $10\times10$. The variance in the data shows how the circuit depths differs for different instances. (b) The probability of staying in the feasible subspace for two different noise levels plotted against the number of QAOA blocks when taking into account the circuit depth from (a).}
\label{fig:withrouting}
\end{figure*}
Most available quantum hardware however has limited connectivity, thus routing of the quantum circuit  becomes necessary. This unavoidably will increase the circuit depth, which in turn increases the amount of noise in the system. To benchmark this effect, we here assume the hardware graph resembling a 2D grid, typical for the prevalent superconducting qubit devices \cite{arute2019quantum}. If the quantum circuit graph cannot be mapped on the hardware graph, SWAP gates have to be used to route the qubits such that all necessary gates can be carried out. With a SWAP network of depth $N$ \cite{kivlichan2018quantum}, it would be possible to allow for all-to-all connections requiring only for a linearly connected hardware graph. However, in the NISQ era, where the circuit depth is a valuable and expensive asset, for circuit graphs with sparse connectivity, it is possible to find SWAP networks with depth smaller than $N$ \cite{cowtan2019qubit}. 

As compiling strongly depends on the problem instance, we generate $100$ random 3-regular graphs with $|V|=30$ vertices, which require $90$ physical qubits to be implemented. As hardware graph, we assume a square grid of size $10\times 10$.
To route the quantum circuit, we use $\mathrm{t|ket\rangle}$ \cite{cowtan2019qubit, sivarajah2020t} and calculate the circuit depth of the routed circuit for various number of QAOA blocks, shown in Fig.~\ref{fig:withrouting}(a). We use the found depth to calculate the probability of staying in the feasible subspace for the noise levels $\eta=\{10^{-3},10^{-4}\}$, resulting in Fig.~\ref{fig:withrouting}(b). In comparison to Fig.~\ref{fig:XY}, we see that the routing drastically worsens the results. 

For both hardware assumptions, we see that already for such problems of rather moderate size, for realistic error rates $10^{-2}$-$10^{-3}$ \cite{arute2019quantum, wright2019benchmarking, barends2014superconducting}, the probabilities are below reasonable experimental sampling requirements of $10^4$-$10^6$ shots \cite{arute2019quantum}. Keeping in mind that QAOA is a hybrid algorithm, independently from the hardware topology many samples will be needed to estimate the cost function up to a precision required for making reasonable parameter updates \cite{mcclean2018barren}.

\subsubsection{Comparison to QAOA with energy penalties}

To stay in the feasible subspace, XY-QAOA uses the mixer in Eq.~(\ref{eq:mixing}). It was proposed as an alternative to X-QAOA that is expected to perform better under no noise, thanks to the design which allows XY-QAOA to restrict exploration to the subspace of valid solutions of the full Hilbert space \cite{wang2020x}.  The alternative, X-QAOA, in analogy with a technique commonly used in quantum annealing, could use penalty terms to encourage the evolution to remain in the feasible subspace while using the standard transverse field mixing operator of X-QAOA $H_\mathrm{X}=\sum_v \sum_j \sigma_x^{(v,j)}$. For Max-$\kappa$-Colorable-Subgraph, the following could serve as a penalty term.
\begin{align}
 H_\mathrm{pen}=\alpha 
    \frac{1}{2}\sum_{v\in E} \left((2-\kappa)\sum_j \sigma_z^{(v,j)} +\sum_{j,j'}\sigma_z^{(v,j)}\sigma_z^{(v,j')}\right)\;. 
    \label{eq:penalty}
\end{align}
At first glance, it seems possible that the transverse field mixer could improve the probability of remaining in the feasible subspace under noise because, even though it takes elements in the feasible subspace to the non-feasible subspace, it can also take elements from the non-feasible subspace to the feasible subspace, which the $XY$ mixer cannot. On the other hand, it behaves like collective bit-flip noise on all qubits, which, in principle, could be corrected, but because of its simultaneous action on all qubits with the same $\beta$ angle, it is very unlikely to rotate all qubits back to the feasible subspace. Hence the analysis of  Sec.~\ref{sec:probability}, particularly Eq.~(\ref{eq:probability_N}) and Fig.\ref{fig:decay}(b) showing that the probability of staying in the feasible subspace decreases monotonically with noise, suggest that the transverse-field mixer causes more harm than good even in the presence of noise.

Here, we give two complementary arguments for why we would expect that the advantage of XY-QAOA over X-QAOA still holds in presence of noise. 
In the first paragraph (a), we motivate that the action of X-mixer does not lead to an increase of the probability to stay in the feasible subspace, at least under depolarizing noise, while in paragraph (b) we show implementing the penalty term in the X-QAOA approach requires a higher circuit depth than XY-QAOA.
\paragraph{Effect of noise}
Let us first compare how different mixers are affected by a local depolarizing channel applied to the feasible subspace projector $\mathcal{P}_\mathrm{feas}$. We select a projector to the entire feasible subspace, instead of a projector to a particular state, because we aim to find an ``average'' behavior, and not restricting the analysis to a special algorithm initialization. Since the feasible subspace projector is a linear combination of product states projectors (e.g. for a 3-qubit system $\mathcal{P}_\mathrm{feas} = P_{\ket{001}}+ P_{\ket{010}} + P_{\ket{100}}$, where $P_{\ket{\psi}}=\ket{\psi}\bra{\psi}$) and as such is invariant under the action of the XY mixer. Therefore, the action of any local noise model, will preserve the linear combination product structure. Local depolarizing noise will additionally preserve the diagonal structure in the computational basis (it suffices to consider action of the noise channel on projectors $\ket{0}\bra{0}$ and $\ket{1}\bra{1}$ which leads to $\rho=\frac{1}{2}\mathbb{1}\pm\frac{\mu}{2}Z$, respectively, with $\mu = 1-\frac{4\eta}{3}$ and $\eta$ being the depolarizing rate). This behavior means, that any probability loss can only be associated with the action of the depolarizing channel. However, for a standard $X$ mixer, we additionally have the dependence on the $\beta$ angle, which can lower the probability of staying in feasible subspace. The XY mixer provides an upper bound for the $X$ mixer's probability of staying in the feasible subspace for level-$1$ QAOA, 
\begin{equation}
    \Tr\Big[P_F\mathcal{E}\big(U_XP_F U_X^\dag\big)\Big]\le \Tr\Big[P_F\mathcal{E}\big(U_{XY}P_F U_{XY}^\dag\big)\Big]\;,
    \label{eq:bound}
\end{equation}
where $U_X\equiv U_X(\beta_x)=\exp[-i\beta_x \sum_i \sigma_x^{(i)}]$, and $U_{XY}\equiv U_{XY}(\beta_{xy})=\exp[-i\beta_{xy} H_{XY}]$, with $H_{XY}$ given in Eq.~\eqref{eq:mixing}, and the inequality holds for an arbitrary pair of  $\beta_x$ and $\beta_{xy}$.

\begin{figure}[t!]
  \includegraphics[width=\linewidth]{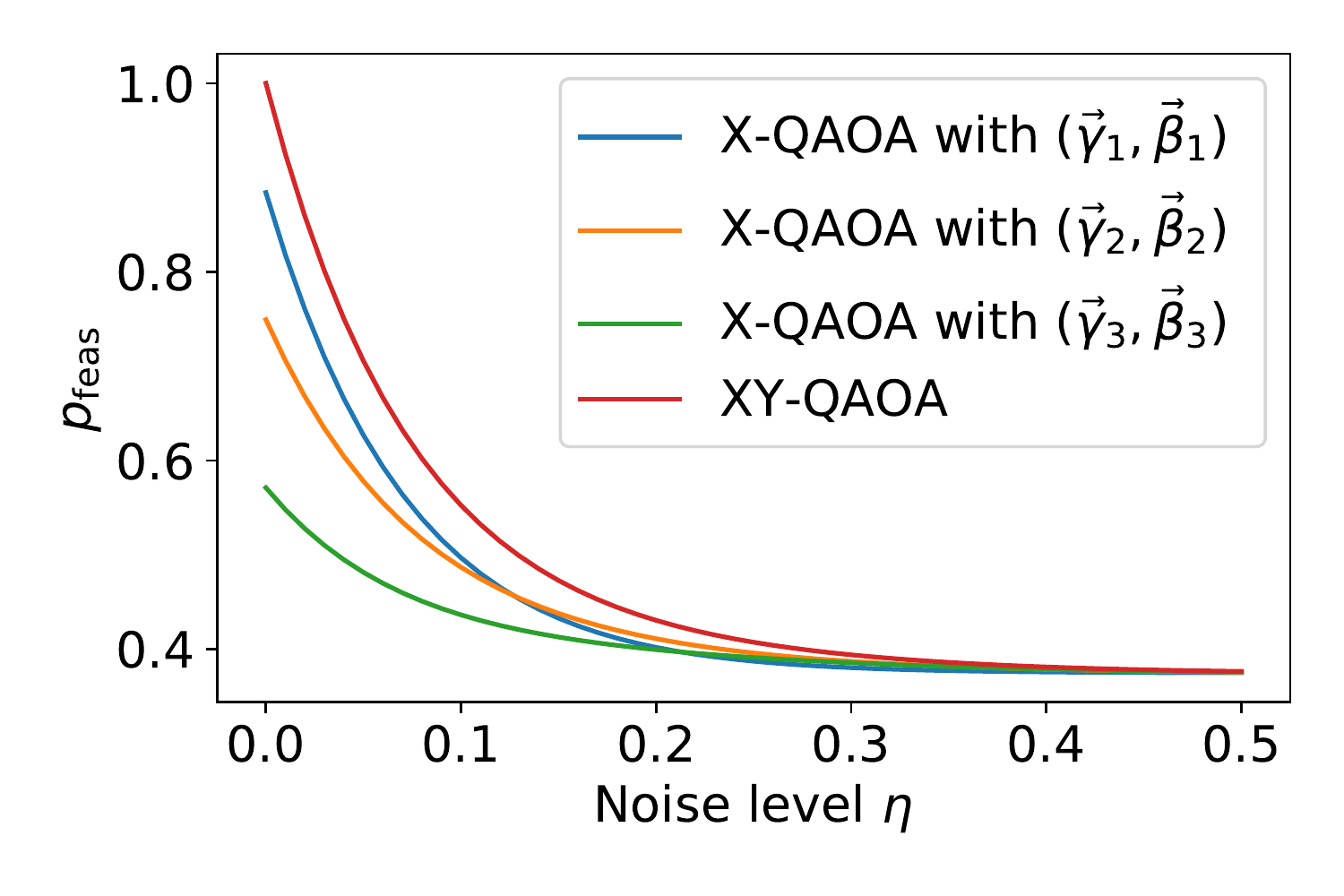}
  \caption{The probability to stay in the $N=1$-subspace of $\kappa=3$ qubits for 4 layers of XY-QAOA and X-QAOA. For XY-QAOA (red), the choice of the parameters does not affect the probability, while for X-QAOA, the following parameters combinations were used:  $\vec{\beta}_1 = (1.5, 0.2, 0.9, 3.8)$ and $\vec{\gamma}_1= (1.5, 3.2, 0.9)$  (blue), $\vec{\beta}_2=(1.5, 1.7, 1.9, 0.8)$ and $\vec{\gamma}_2=(0.4, 1.2, 1.7)$  (orange), $\vec{\beta}_3 = (2.2, 3.7, 2.9, 2.8)$ and $\vec{\gamma}_3=(1.4, 0.9, 3.1)$ (green). 
  Parameters are given in the form $\vec{\beta}=(\beta_1,\beta_2,\beta_3,\beta_4)$ and $\vec{\gamma}=(\gamma_2,\gamma_3,\gamma_4)$. The angles were selected such that the sum of all $\beta$ is close to a multiple of $2\pi$; if selected differently (e.g. the sum close to the odd multiples of $\pi$), one would observe even worse performance. Since the first application of the penalty Hamiltonian (see Eq.~\ref{eq:penalty}) does not affect the probability, $\gamma_1$ is omitted. Our investigation suggests that the selected angles represent the general behavior that is in accordance with Eq.~(\ref{eq:bound}).}
  \label{fig:XY_vs_X}
\end{figure}

Any rotation out of computational basis plane will decrease the probability of staying in the feasible subspace. Eq.~(\ref{eq:bound}) technically only holds for one level of X and XY mixers. However, numerical results on $4$ layers of mixers shown in Fig.~\ref{fig:XY_vs_X} strengthen our argument that this bound also holds for repeated action of the mixers. In the numerical analysis we selected a set of angles, that is representative of the overall performance, therefore we expect similar behavior for any other set of angles.

\paragraph{Circuit depth}

Furthermore, both approaches  require the implementation of the phase separation unitary $U_\mathrm{P}$. The unitary generated by the penalty Hamiltonian, see Eq.~(\ref{eq:penalty}), requires two-qubit gates between all pairs of qubits representing a logical vertex. The same holds for the action of the mixing unitary, see Sec.~\ref{sec:xyqaoa-fully-connected}. Thus, on a fully-connected quantum computer the mixing unitary of XY-QAOA requires the same depth as the implementation of the penalty term of X-QAOA.  However, X-QAOA requires an additional layer of local $X$ rotation yielding a higher total circuit depth for X-QAOA. The depth on specific hardware will depend on which two-qubit gate or gates are natively implemented.

\section{Mitigating the effect of noise}
\label{sec:embeddingsanderrormitigation}

In the previous section, we have used states with a single particle of $\kappa$ qubits to encode $\kappa$ colors and found that the effect of noise makes it intractable to push forward to problems with hundreds of variables with noise levels present in currently available quantum hardware. To make XY-QAOA ready for applications in the NISQ era, one either has to substantially lower the noise levels, find noise-resilient ways to encode the problems or develop error mitigation techniques. In the following, we discuss the two latter options.

In Sec.~\ref{sec:large_subspace} we consider encoding the feasible states into a subspace corresponding to a larger particle number, which typically is a larger subspace than the $N=1$ subspace, and examine how this encoding affects the probability of staying in the feasible subspace.
In Sec.~\ref{sec:correction}, we introduce a bit-flip quantum error correction code adapted to specifically preserving a particle number preserving subspace and evaluate its efficiency. This code demonstrates the potential gains in resource efficiency through symmetry-aware quantum error correction techniques.

\subsection{Encoding in larger particle number subspaces}
\label{sec:large_subspace}

Utilizing a larger fraction of the full Hilbert space as feasible subspace increases the number of valid solutions over unwanted solutions. 

On the other hand, encoding in a different subspace will also change the expression of the cost function, incurring a different circuit depth and therewith changing the amount of noise in the computation.

In this section, we explore if such an approach can be helpful to uphold the probability of sampling valid solutions.
We again focus on the Max-$\kappa$-Colorable-Subgraph problem. The cost function given in Eq.~$\eqref{eq:cost_one_hot}$, can be easily adapted to other particle number subspaces by including higher-order interactions. The particle number $N$ states of $\nq$ qubits can be used to implement the Max-$\binom{\nq}{N}$-Colorable-Subgraph problem with cost function
\begin{align}
    f_\mathrm{C}^\mathrm{N}
    & = \sum_{(v,v')\in E}\frac{1}{N!}
    \sum_{
    \substack{i_1,\dots,i_\mathrm{N}\\ i_1\neq\dots\neq i_\mathrm{N}}}^{\nq}x_{v,i_1}x_{v',i_1}\cdots x_{v,i_\mathrm{N}}x_{v',i_\mathrm{N}}
    \;,
\end{align}
which includes up to $2\mathrm{N}$-local terms.

As example, for $\kappa=6$ colors, we could, as before, use the 6 states with single particle of 6 qubits. Alternatively the 6 states with two particles of 4 qubits could be used. While the first approach uses a fraction of $6/2^6\approx 9\%$ of the respective full Hilbert space, the latter approach uses $6/2^4\approx 38\%$ of all states. 
To see whether this gain in fraction yields also a gain in robustness against noise, we have to calculate the circuit depth required to implement the different cost function.

The cost function with $N=2$ particles reads
\begin{align}
    f_\mathrm{C}^{N=2}=\sum_{(v,v')\in E}\frac{1}{2}\sum_{\substack{i,j\\i\neq j}}^{\nq=4} x_{v,i}x_{v,j}x_{v',i}x_{v',j}\,,
    \label{eq:costfunction_6_colors_H_2}
\end{align}
and includes a product of 4 binary variables. To construct the corresponding problem Hamiltonian we again replace the binary variables $x_{a,b}$ with Pauli operators, which results in a Hamiltonian $H_\mathrm{P}^{N=2}$ with up to $4$-local terms. 
We recognize that not all terms in the Hamiltonian are needed to represent the problem faithfully. For example, as in Sec.~\ref{sec:performance_under_noise}, the sum over the 1-local terms on each vertex is proportional to the particle number operator $N_v$ of that vertex in the feasible subspace, 
\begin{align}
    \tilde{N}_v=\sum_i\sigma_z^{(v,i)}=2N_v-4=0\,,
\end{align}
and thus not needed to encode the problem faithfully. To make the circuit as short as possible, we next show how to reduce the number of terms in the Hamiltonian, when restricted to the feasible subspace. 
For example, the 2-local terms of the form $\sum\sigma_z^{(v,i)}\sigma_z^{(v',j)}$ can be simplified in the following way:
\begin{align}
    \label{eq:2-local-reduction}
    \sum_{\substack{i,j\\i\neq j}}\sigma_z^{(v,i)}\sigma_z^{(v',j)}&=\sum_i\sigma_z^{(v,i)}\sum_j\sigma_z^{(v',j)}- \sum_i\sigma_z^{(v,i)}\sigma_z^{(v',i)}\nonumber\\
    &=\tilde{N}_v \tilde{N}_{v'}- \sum_i\sigma_z^{(v,i)}\sigma_z^{(v',i)}\nonumber\\
    &=- \sum_i\sigma_z^{(v,i)}\sigma_z^{(v',i)}
\end{align}
We thus only have to consider the second term, which is a sum of 2-local interactions between the same qubits of vertex $v$ and $v'$. For the other 2-local terms, the calculation is analogous. Similarly, we can show that all 3-local terms are vanishing, 
\begin{align}
    \label{eq:3-local-reduction}
    &\phantom{=} \sum_{\substack{i,j\\i\neq j}} \sigma_z^{(v,j)}\sigma_z^{(v,j)}\sigma_z^{(v',i)}\nonumber\\
    &= \sum_{i}\sigma_z^{(v,j)}\sum_i \sigma_z^{(v,i)}\sigma_z^{(v',i)}-\sum_{i}\sigma_z^{(v,i)} \sigma_z^{(v,i)}\sigma_z^{(v',i)}\nonumber\\
    &= \tilde{N}_v\sum_i \sigma_z^{(v,i)}\sigma_z^{(v',i)}-\sum_{i}\sigma_z^{(v',i)}\nonumber\\
    &= \tilde{N}_v\sum_i \sigma_z^{(v,i)}\sigma_z^{(v',i)}-\tilde{N}_{v'}=0
\end{align}
For the 4-local terms, we cannot easily extract the particle number operator, but when restricted to the feasible subspace, we recognize that only 4 terms are necessary to represent the sum distinctly. Careful collection of all remaining terms yield the Hamiltonian
\begin{align}
    H_\mathrm{P}^{N=2}&=\sum_{(v,v')\in E} H_{(v,v')}\hspace{5mm}\mathrm{with}\nonumber \\
   H_{(v,v')} &=\frac{1}{4}\sum_{j=1}^4 \sigma_z^{(v,j)} \sigma_z^{(v',j)}\nonumber\\
    &\phantom{=}+ 
    \frac{1}{8}\Big( 
    \sigma_z^{(v,1)} \sigma_z^{(v,2)} \sigma_z^{(v',1)} \sigma_z^{(v',2)}\nonumber\\
    &\phantom{=}+ \sigma_z^{(v,3)} \sigma_z^{(v,4)} \sigma_z^{(v',3)} \sigma_z^{(v',4)}\nonumber\\
    &\phantom{=}+ 2\sigma_z^{(v,1)} \sigma_z^{(v,4)} \sigma_z^{(v',1)} \sigma_z^{(v',4)}\nonumber\\
    \label{eq:simplified_N_2}
    &\phantom{=}+ 2\sigma_z^{(v,2)} \sigma_z^{(v,4)} \sigma_z^{(v',2)} \sigma_z^{(v',4)}\Big)\;,
\end{align}
where the last expression for 4-qubit interactions is invariant under qubit permutation for each node. For each connection $(v,v')\in E$ in the problem graph $G$, we thus have $4$ two- and $4$ four-qubit interactions. 
The two-qubit interactions can be executed in one layer. The 4-qubit interactions, cf. Eq.~\eqref{eq:simplified_N_2}, can be executed in 3 layers, assuming that we can carry out any 2- or 4-qubit gates as a single gate, i.e. without decomposing them into native 1- and 2-qubit gates.  For a $k$-regular graph, the problem Hamiltonian then requires depth $d_\mathrm{P}^{N=2}=4(k+1)$ to be implemented. 
Independent from the cost function, the same version of the mixing Hamiltonian, cf. Eq.~(\ref{eq:mixing}), can be used as it preserves any particle number and in analogy to before, results in a circuit of depth $d_\mathrm{XY}^{N=2}=4$.

\begin{figure}[t!]
    \centering
    \includegraphics[width=1\linewidth]{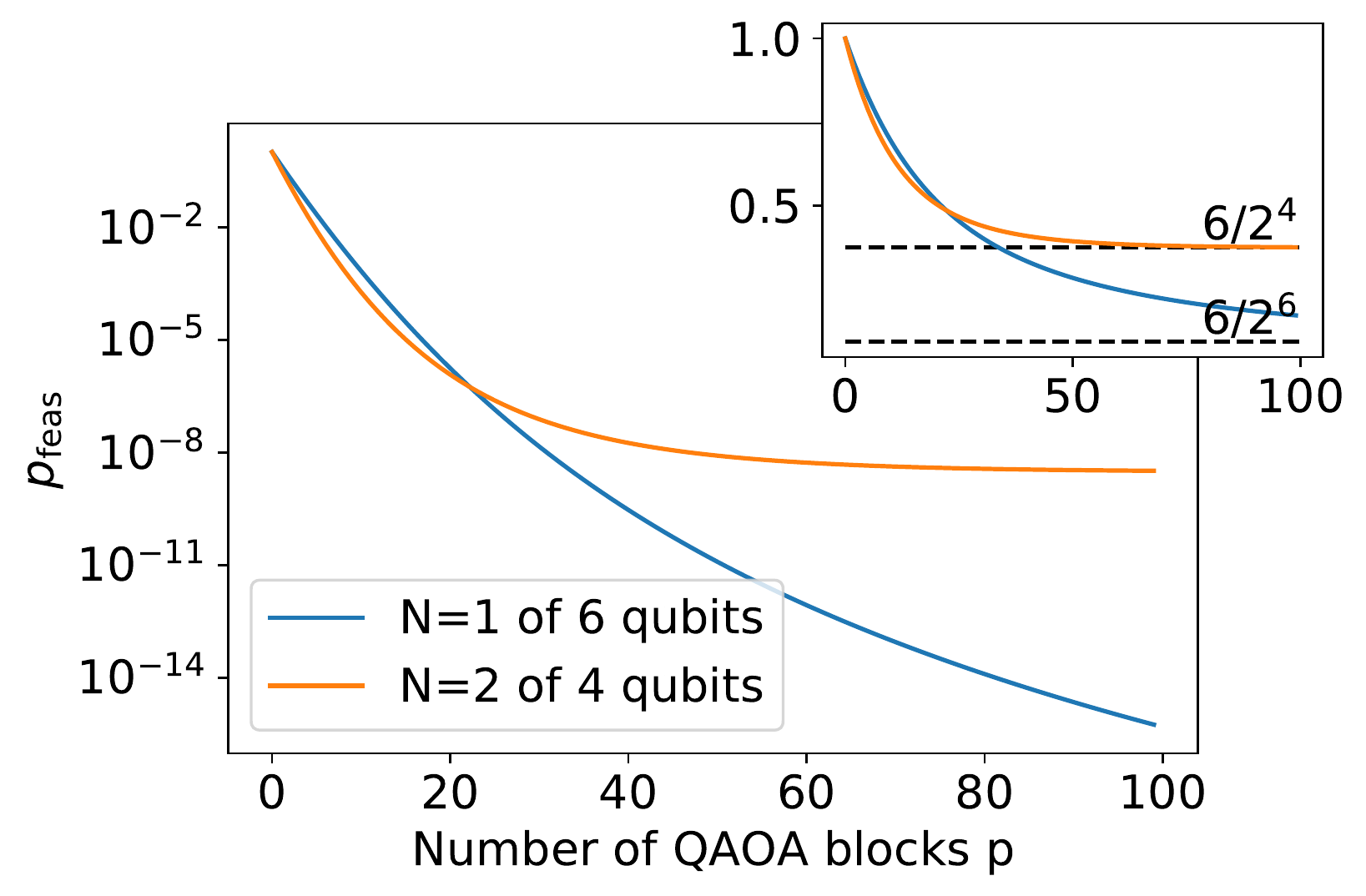}

  \caption{Comparison of the probability to stay in the feasible subspace of $N=1$ states of $6$ qubits and $N=2$ states of $4$ qubits for a system size of $|V|=20$ (main plot) and $|V|=1$ (inset) logical vertices with a noise level of $\eta=10^{-3}$. }
  \label{fig:comparison_encodings}
\end{figure}

In Fig.~\ref{fig:comparison_encodings}, we show the probability of staying in the feasible subspaces for both encodings as function of the number of QAOA blocks. As clearly visible in the plot, in the regime of interest, i.e., where $p_\mathrm{feas}$ is fairly high,
the overhead of implementing the more complex problem Hamiltonian obliterates any possible advantage of the larger subspace. For very deep circuits, the higher particle number shows advantage. However, at this level, 
due to the low value of $p_\mathrm{feas}$,
we would mostly see non-valid samples.  We note again that these numbers would look even worse in a real experiment, as we would have to account for decomposing the gates into a native gate set and for the routing of the circuit to a restricted hardware graph. As the encoding into the higher particle number subspace results in a higher gate count with more complex gates, taking these effects into account would make the difference in both encodings even more visible.
However, this first result highlights that the way of encoding the problem has significant influence on the robustness of the quantum algorithm. Whether schemes with higher-order terms pose an alternative to the one-hot encoding also depends on the availability of complex many-qubits gates in the next generations of NISQ devices.

\subsection{Correction of symmetry-breaking errors}
\label{sec:correction}

\begin{figure*}[!htb]

\minipage{0.5\linewidth}

\includegraphics[width=\linewidth]{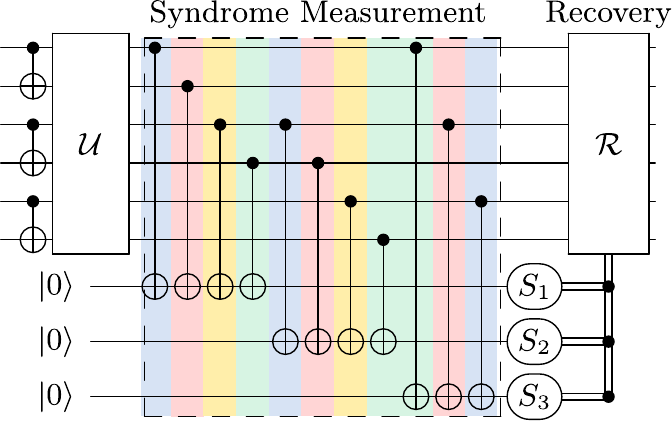}\llap{
  \parbox[b]{6.0in}{(a)\\\rule{0ex}{2.2in}
  }}
\endminipage\hfill
\minipage{0.5\linewidth}
\rlap{
  \parbox[b]{1.1in}{(b)\\\rule{0ex}{0.5in}
  }}
\begin{tabularx}{0.5\linewidth}{cccc}
\toprule
    $\braket{S_1}$ & $\braket{S_2}$  & $\braket{S_3}$  & Recovery\\
    \hline
    1 & 1 & -1 & $X_3$                   \\
    1 & 1 & 1  & $X_4$                    \\
    1 & -1  & -1 & $X_1$                    \\
    1 & -1  & 1  & $X_2$                    \\
    -1  & 1 & -1 & $X_5$                     \\
    -1  & 1 & 1  & $X_6$                    \\
    -1  & -1  & 1  & $\mathbb{1}$ \\
    \hline
    -1  & -1  & 1 & N/D
    \vspace{7mm}
    \end{tabularx}
\endminipage\hfill

\minipage{0.5\linewidth}

\includegraphics[width=\linewidth]{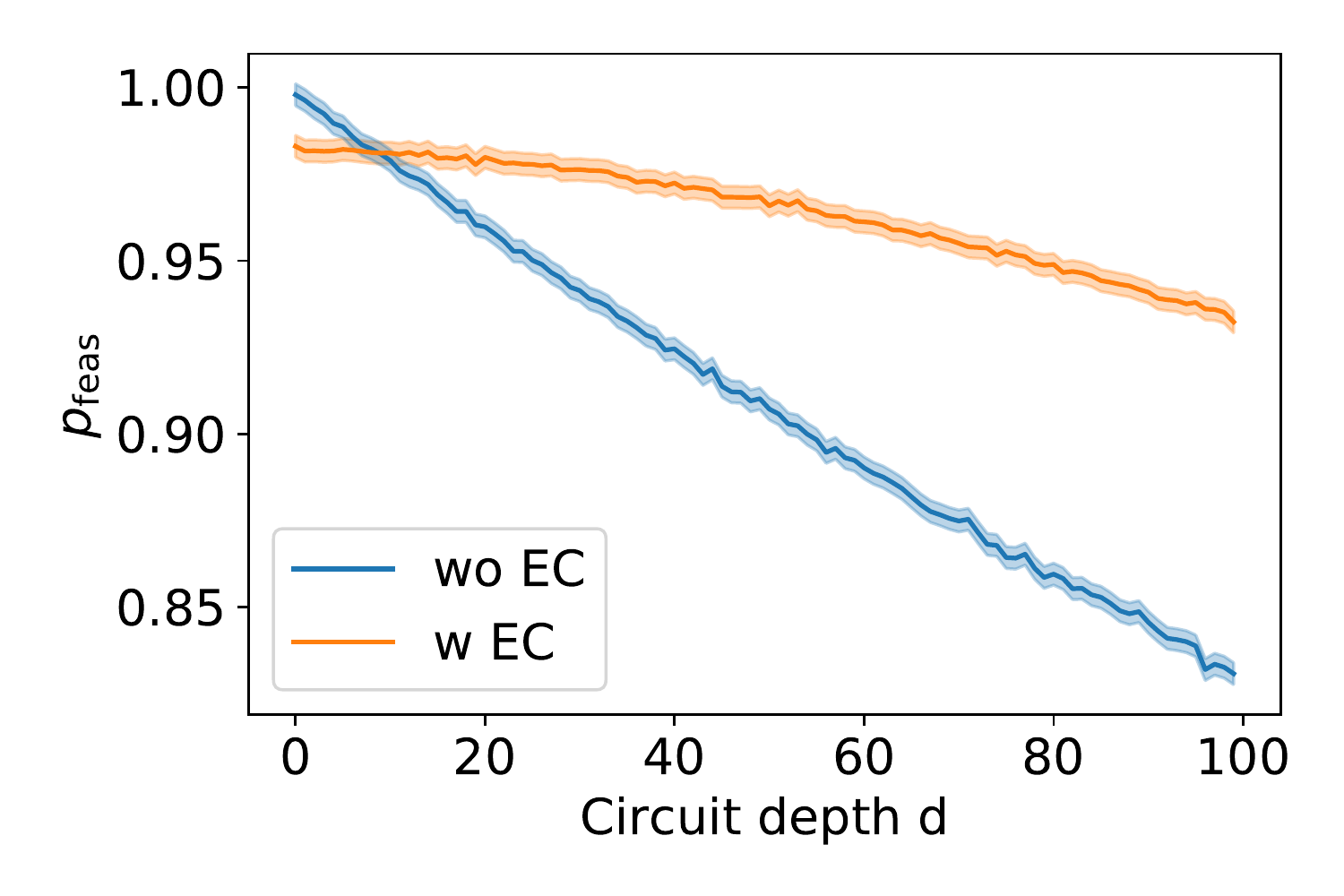}\llap{
  \parbox[b]{6.0in}{(c)\\\rule{0ex}{2.2in}
  }}
\endminipage\hfill
\minipage{0.5\linewidth}

\includegraphics[width=\linewidth]{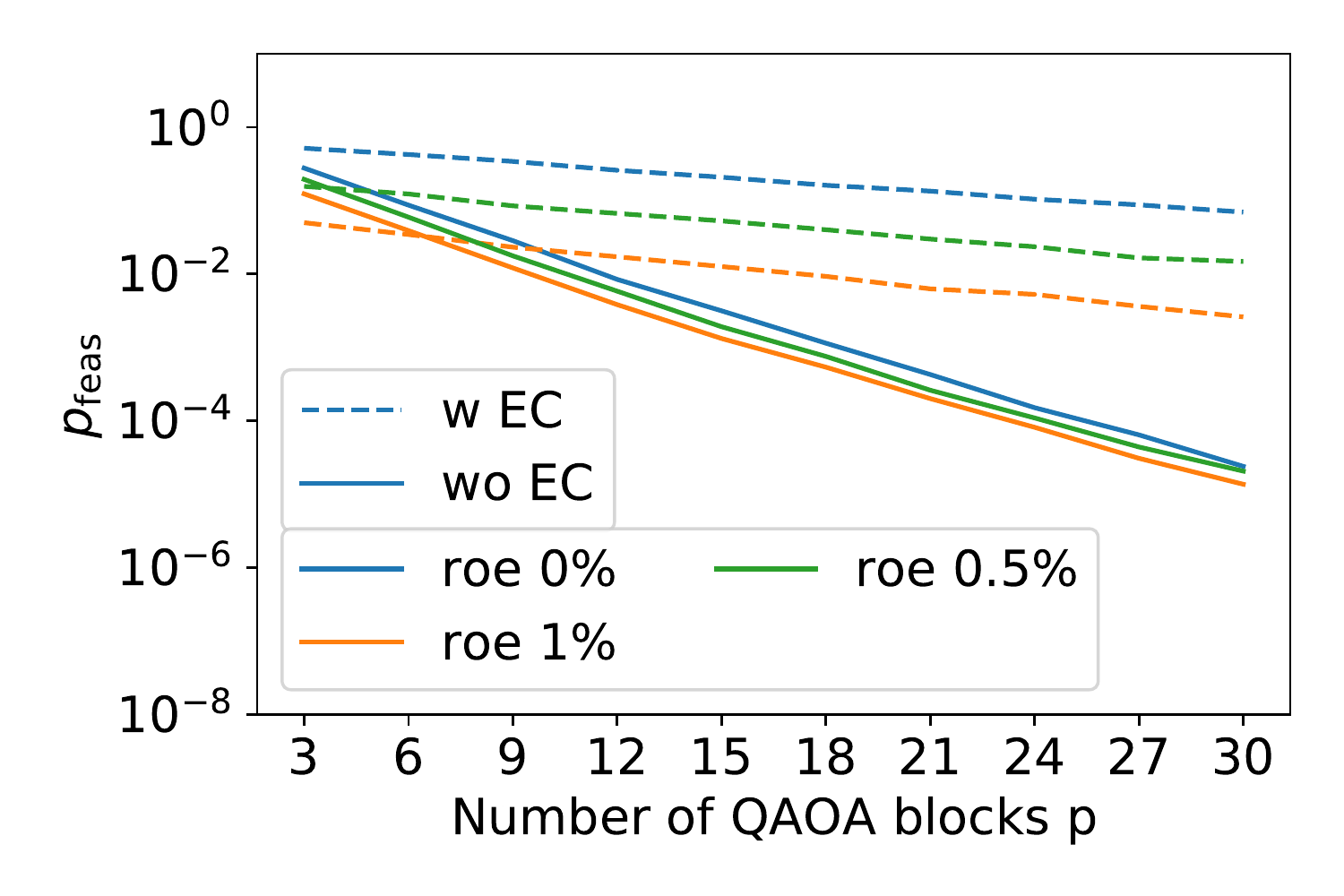}\llap{
  \parbox[b]{6.0in}{(d)\\\rule{0ex}{2.2in}
  }}
\endminipage\hfill
\caption{(a) Error correction circuit correcting for a single bit flip error on the data qubits with the syndromes given in Eq.~(\ref{eq:syndromes}). The colors highlight the gates which can be carried out in one layer on a fully-connected quantum computer. According to the syndrome measurements, one of the recovery actions given in (b) is carried out where $X_i$ defines the local bit flip operation $\sigma_x^{(i)}$ on qubit $i$. (c) The probability to sample a valid state for a noise level of $\eta=10^{-3}$  when applying the error correction circuit after $d$ layers at the end of the computation - in comparison to no error correction. For this plot, we simulated the circuit execution and sampled $10^5$ times from the output state for each circuit depth. The shaded area reflects the statistical inaccuracy. (d) Comparison of the error correction scheme for XY-QAOA applied to the Max-$3$-Colorable-Subgraph problem of a 3-regular graph with $|V|=30$ vertices. For this plot, a noise level of $\eta=10^{-3}$ and readout errors (roe) $\{0\%,0.5\%,1\%\}$ were used. As in (c), $10^5$ samples where used.}
\label{fig:error_mitigation_circuit}
\end{figure*}

In the previous section we have analyzed the probability of sampling valid solutions when encoding the Max-$\kappa$-Colorable-Subgraph problem in a higher particle number subspace than the $N=1$ particle number subspace. Due to the overhead in the circuit depth when implementing the different cost function, we did not encounter an advantage over the common one-hot encoding. 
In this section, we discuss a possibility to correct the symmetry-breaking errors for the example of encoding $\kappa=3$ colors into the $N=1$ subspace.  We design an adaption of the bit flip code  \cite{nielsen2002quantum, rieffel2011quantum} that requires fewer ancillary qubits by exploiting the system's particle number symmetry, as shown in Fig.~\ref{fig:error_mitigation_circuit}(a),
and analyze its effect.
The reduction in resources is possible because we aim only to correct states in the $N=1$ particle number subspace.

In the original bit flip code, two copies of each qubit are needed to be able to identify a single bit flip through a majority vote on the three qubits.  The present adaption benefits from the additional information on the LPNC of the data qubits, which reduces the number of necessary copies from two to one. 
The syndromes
\begin{align}
    S_1&=Z_1Z_2Z_3Z_4\mathbb{1}_5\mathbb{1}_6\,,\nonumber\\
    S_2&=\mathbb{1}_1\mathbb{1}_2Z_3Z_4Z_5Z_6\,,\nonumber\\
    S_3&=Z_1\mathbb{1}_2Z_3\mathbb{1}_4Z_5\mathbb{1}_6\,,
\label{eq:syndromes}
\end{align}
with the subscripts $\{1,3,5\}$ and $\{2,4,6\}$ labeling the original qubits and their copies respectively, are able to detect a single bit flip error on the data qubits. To readout the syndromes non-invasively, i.e. without destroying the quantum state, three additional ancilla qubits are needed. As in all quantum error correction schemes, measurements of the ancillas do not provide us with any additional information about the data qubits' quantum state. After measuring the ancillas, the recovery actions, summarized in Fig.~\ref{fig:error_mitigation_circuit}(b), correct the error by applying a local bit flip operation. The effectiveness of this error correction procedure depends on the probability of errors in the input state,
the noise in the correction circuit itself and the readout error of the ancillas. We note that this kind of error correction scheme does not correct for errors inside the feasible subspace. 

To discuss how this scheme performs under realistic noise levels, we have to decide how often we want to apply it during the quantum computation. Under the assumption that the error correction circuit is perfect and noiseless, we could apply it after each layer of the circuit and would always see an improvement of the results. However, in realistic scenarios, the error correction circuit itself is subject to noise.  To analyze how often we can apply the scheme without worsening the results, we look at the following setup: we initialize the system in a feasible state, subsequently apply $d$ layers of QAOA with noise and afterwards apply the syndrome circuit with the ancilla qubits initialized in the $\ket{0}$ state. We use the same noise model as before and apply noise channels on all $9$ qubits after each layer. On a fully-connected quantum computer, the correction circuit can be carried out in depth $4$, highlighted by the colors in Fig.~\ref{fig:error_mitigation_circuit}(a). Based on the measurement outcomes of the ancillas, we then apply the recovery action on the data qubits and sample from the final state. In Fig.~\ref{fig:error_mitigation_circuit}(c) we show the probability to sample a feasible state in dependence of the circuit depth $d$ prior to the onset of correction, in comparison with the situation of no error correction involving only $3$ qubits. As expected, if the input state has no or little noise, the correction scheme even worsens the result. However, for noisier input data, we see a clear advantage in applying the error correction code.

In Sec.~\ref{sec:xyqaoa-fully-connected}, we applied XY-QAOA to the Max-$3$-Colorable-Subgraph problem of a 3-regular graph and saw that the results are quickly beyond reasonable sampling assumptions. We here analyze how these results change when we apply the error correction scheme after every 3 QAOA blocks (i.e. 21 layers, cf. Sec.~\ref{sec:xyqaoa-fully-connected}) of the XY-QAOA circuit. To make the situation even more realistic to current hardware prototypes, we include readout errors on all qubits by defining the symmetric probabilities $P(0|1)=P(1|0)=\mathrm{roe}$ that a qubits' state is readout falsely. For current technologies, readout errors are between $0.1\%$ and $5\%$ \cite{arute2019supplementary, myerson2008high}. Such errors lead to incorrect recovery actions, tempering the effectiveness of the correction. 
In Fig.~\ref{fig:error_mitigation_circuit}(d), we show the probability to find a feasible state for a graph with $|V|=30$ vertices in dependence of the number of QAOA blocks for the situations with and without error correction, a noise level of $\eta=10^{-3}$ and readout errors (roe) $\{0\%,0.5\%,1\%\}$. As clearly visible, beyond the crossover, error correction improves the results by several orders of magnitude. As expected, the read out errors have a stronger effect on the computation when error correction is applied. However, we still see a clear advantage in applying this simple error correction scheme for all studied readout error strengths.

In this example, we have assumed a fully-connected quantum computer. On devices with limited qubit connectivity, the overhead of routing the error correction has to be taken into account. 
Moreover, the recovery operation requires precise measurements and fast classical control loops, which, while challenging on certain hardware prototypes where measurement takes a long time, on some other hardware prototypes is expected to be readily achievable \cite{andersen2019repeated}.

\section{Conclusion \& Outlook}
\label{sec:conclusion}
In this paper, we have studied the robustness of quantum algorithms with local particle number conservation under the influence of noise. We found an exact combinatorial expression to calculate the probability of staying in particle number subspaces under the influence of local depolarizing noise. With these findings we benchmarked the robustness of XY-QAOA applied to the Max-$\kappa$-Colorable-Subgraph problem and analyzed the influence of the choice of the problem encoding on the robustness of the algorithm. We moreover discussed the possibility of correcting for symmetry-breaking errors. Our results highlight the importance of finding novel and clever encoding techniques or error mitigation strategies to make noisy small-scale devices applicable to solving optimization problems with constraints.

In this work, our main emphasis focused on question how likely it is to see a feasible sample at the end of a XY-QAOA, which also upper bounds the performance of the algorithm. However, the present analysis could also be applied to other quantum algorithms with particle number symmetry, e.g. quantum chemistry applications.
In variational quantum eigensolver (VQE) or state preparation algorithms applied to many-body electronic systems, 
among other symmetries, often it is desired that the number of electrons is preserved.
Ways to enforce such symmetries, just like in QAOA, include adding terms in the VQE energy function that penalize symmetry violations \cite{McClean16, Ryabinkin19} or designing quantum circuits to limit the quantum evolution within the subspace spanned by states with the desired symmetries regardless of the values of the variational parameters \cite{Seki20,Gard20}.  The latter in noiseless case often predicts superior results compared with the former.  Our noise analysis is based on preserving the total $S^z$ of a spin (qubit) system, while the encoding of fermionic system into qubits is impeding direct application of the analysis we performed to electronic systems, we believe the conclusion in high-level applies to all such designed algorithms.

Moreover, these results could be generalized from particle number symmetries to more general symmetries. 
Another application beyond benchmarking quantum algorithms lies in using particle number subspaces to benchmark quantum hardware. 

\section*{ACKNOWLEDGMENTS}
The authors would like to thank Andrea Skolik, Michael J. Hartmann, Jason M. Dominy, Sheir Yarkoni and all members of the QuAIL team for helpful discussions. MS was supported by the USRA Feynman Quantum Academy funded by the NAMS R\&D Student Program at NASA Ames Research Center and by the Air Force Research Laboratory (AFRL), NYSTECUSRA Contract (FA8750-19-3-6101), and thanks Saba Hussain and Zeal Panchal for organization. MS and ML thank André Radon and Florian Neukart for support. EGR, MS, FW, and ZW are grateful for support from NASA Ames Research Center,  the AFRL Information Directorate under grant F4HBKC4162G001, and DARPA under IAA 8839, Annex 114. This project has received funding from the European Union’s Horizon 2020 research and innovation programme under the Grant Agreement No. 828826.

\bibliographystyle{unsrt}
\bibliography{references}
\end{document}